\documentclass{IOS-Book-Article}

\usepackage{mathptmx}
\usepackage{soul}\setuldepth{article}
\def\hb{\hbox to 11.5 cm{}}

\begin{document}

\pagestyle{headings}
\def\thepage{}
\begin{frontmatter}              

\title{Who is responsible? Social Identity, Robot Errors and Blame Attribution}


\author[A]{\fnms{Samantha} \snm{Stedtler}\orcid{0009-0002-4410-5595}%
\thanks{Corresponding Author: Samantha Stedtler, Department of Philosophy, Lund University, Sweden; Email: samantha.stedtler@lucs.lu.se}}
\author[A]{\fnms{Marianna} \snm{Leventi}\orcid{0000-0001-6454-0362}}
\address[A]{Department of Philosophy, Lund University, Sweden}

\begin{abstract}
This paper argues that conventional blame practices fall short of capturing the complexity of moral experiences, neglecting power dynamics and discriminatory social practices. 
It is evident that robots, embodying roles linked to specific social groups, pose a risk of reinforcing stereotypes of how these groups behave or should behave, so they set a normative and descriptive standard.  
In addition, we argue that faulty robots might create expectations of who is supposed to compensate and repair after their errors, where social groups that are already disadvantaged might be blamed disproportionately if they do not act according to their ascribed roles. 
This theoretical and empirical gap becomes even more urgent to address as there have been indications of potential carryover effects from Human-Robot Interactions (HRI) to Human-Human Interactions (HHI). We therefore urge roboticists and designers to stay in an ongoing conversation about how social traits are conceptualised and implemented in this technology. We also argue that one solution could be to 'embrace the glitch' and to focus on constructively disrupting practices instead of prioritizing efficiency and smoothness of interaction above everything else.
Apart from considering ethical aspects in the design phase of social robots, we see our analysis as a call for more research on the consequences of robot stereotyping and blame attribution. 
\end{abstract}

\end{frontmatter}

\section{Introduction}

In this paper, we will argue that our common practices around blame can not capture all aspects of our moral experiences since they do not account for power dynamics and discriminatory social practices. Moreover, we will discuss how this issue is emphasized in interactions with social robots and in the way social identity and norms are imagined through robots.
Errors in Human-Robot interaction (HRI) are common and inevitable due to various reasons that can apply to any developing technology. Companies and researchers are therefore investigating how to rectify these situations, often by testing strategies such as delivering explanations and apologies \cite{mahmood2022owning,lee2010gracefully}. At the same time, some robots are occupying positions that contain traits associated with specific groups of people. For instance, many voice assistants are equipped with female voices and execute feminized domestic labor.\footnote[2]{This has been pointed out by a number of scholars, e.g. Strengers and Kennedy\cite{strengers2021smart}, Rhee\cite{rhee2018robotic}, Perugia and Lisy\cite{perugia2023robot}.}
Since many social roles are grounded in a society organized through oppressive and discriminatory structures, there seems to be a need to consider how people react differently to errors depending on the assumed social identity of the robot.\footnote[3]{For an excellent overview see \cite{oshana2018social}.} In addition, it can be examined whether and how repair strategies should be used in a way to not perpetuate stereotypes. When an error occurs, people usually try to find the cause of it i.e. find out who is responsible, and thus, who is to blame.\footnote[4]{For the purpose of this paper, we will not distinguish between moral responsibility and blame. For theories that distinguish between the two concepts see \cite{mckenna2012conversation} and \cite{zimmerman1988essay}.} Responsibility and blame attribution often go hand in hand with emotions, in the Strawsonian tradition known as reactive attitudes. This connection should not be surprising, since, according to Wilson, “Cybernetics and AI have always been deeply engaged with affect” (\cite{rhee2018robotic} p.57).
The aptness of blaming emotions in these cases has often been discussed, however without necessarily considering power imbalances and gender stereotypes that people are holding.
\newline With the increase of character-holding technology, such as gendered AI systems, there is a risk of extending these discriminatory practices through those systems. For instance, one might build up specific expectations towards people presumably belonging to a certain social category, which can be encouraged by a robot behaving in a way that fits into the specific stereotypes and anticipations that people expect from human groups that they view as subordinates.
This can end in a circular process, where expectations that are encouraged through these encounters with robots might then lead to aggregated problematic behaviors toward humans, and to a normalization of these.
E.g., Mahmood and Huang \cite{mahmood2023gender} compared voice gender, error mitigation strategies, and participant gender. They found that male participants preferred female apologetic voice assistants over apologetic masculine ones and interrupted assistants more often than female participants. These behaviors might pose a risk, since there has been evidence of carryover effects from Human-Robot Interactions (HRI) to Human-Human Interactions (HHI), meaning that the way that people interacted with robots influenced the way they subsequently interacted with humans \cite{erel2022carryover}.

\subsection{Aim and Positionality}
The overall motivation for this paper is to explore whether and how robot gendering can lead to stereotypes and negative reactive attitudes that are extended to humans that fit into the supposed social category.\footnote[5]{We rely on Haslanger's \cite{haslanger2006good} notion of gender as a social category.}
This is in line with frequent calls for placing more importance on studying the effects of HRI on HHI. 
Another goal is to connect this issue to the responsibility literature within moral philosophy, and in particular, Strawson’s reactive attitudes, and look into how this paradigm relates to gender inequalities and the social identities of robots. More specifically, we want to show how one of the most commonly used approaches to themes of responsibility entails flaws that might extend into our interactions with robots. 
Leaning into Vallor's \cite{vallor2024ai} 'AI as a Mirror' metaphor, we argue that instead of threatening functioning moral practices, robots make more apparent what does not work about those practices. To take it one step further, they risk getting us stuck in the past.
To address these issues, we draw from our combined backgrounds in moral philosophy, cognitive science, psychology, and HRI.

\section{Errors and Blame in Human-Robot Interaction}


\subsection{Why Blame?}
Blame plays an important role in upholding moral standards in society by exercising social control and holding others accountable.
Individuals regularly attempt to identify causal relationships between events e.g. finding the root of an error \cite{matsui2021blame,heider2013psychology}. These attributions can influence their following demeanor \cite{groom2010critic}. 
Blaming describes the process of holding responsible for whatever is seen as the cause of a negative outcome \cite{lane2000moral,shaver2012attribution}. Although this definition sounds plausible, there are different understandings of blame depending on the philosophical tradition. Blame is the focus of many discussions about moral responsibility, however for the needs of this paper, we emphasize the strawsonian paradigm. According to a broadly construed strawsonian view our interpersonal relationships should be in the focus of our responsibility investigation, and moral sentiments (or reactive attitudes) play a major role in how we morally interact with each other. Many philosophers suggest that we express our blaming through these reactive attitudes, and that these attitudes express a demand we have towards the person we blame \cite{darwall2009second,wallace1994responsibility,walker2007moral}.
\newline Within social psychology, blame has for instance been assumed to have evolved as a tool for controlling others behavior and to enforce cooperation (socially-regulated blame perspective) \cite{monroe2019people,guala2012reciprocity}. By imposing costs on both the blamer and the violator, blame is usually constrained by the need for evidence justifying one's moral judgment. However, there seem to be various cognitive biases which are shaping the allocation of blame and responsibility. One example of this is an asymmetry in blame and praise judgments, meaning that responsibility for positive events tends to be assigned to a larger group of people while responsibility for negative events is assigned to fewer people \cite{schein2020praise}. One explanation for this could be the fact that praise is less costly than blame, where blaming wrongfully might lead to a high social cost \cite{malle2014theory}. Moreover, being provided with information regarding an outcome, people are unable to ignore that information \cite{janoff1985cognitive}; instead, they overestimate the predictability of the outcome (also referred to as creeping determinism or hindsight bias) and causal links between events, resulting in victim-blaming. The latter has previously also been explained by referring to the need to believe in a just and controllable world \cite{lerner1980just,wortman1976attributions}.
Consistent with that is that, while Monroe and Malle \cite{monroe2019people} found that participants updated blame judgments in the face of new evidence, biases could arise when the social requirement for warrant was loosened, as seen when evaluating outgroup members. In general, blame practices seem like an appropriate start to investigate our moral practices and social reactions and understandings.


\subsection{Perception of Responsibility in Cases of Robot Error}

It has been shown that people attribute mistakes to computers and that agents, as well as robots who are assumed to be more autonomous, are seen as more responsible than less autonomous agents \cite{friedman1995s,serenko2007interface}. Autonomy and agency appear to influence responsibility attribution in ways that will be shown to connect to norms and social identities.\footnote[6]{For a discussion of these concepts in relation to sexism and misogyny see \cite{manne2017down}.} Agency, i.e. being capable of regulating acting and reasoning \cite{matsui2021blame,gray2007dimensions}, has been negatively correlated with attribution of responsibility. For instance, \cite{miyake2019mind} found that participants attributed more responsibility to interaction partners (human, computer, and robot) with less perceived agency. This is in contrast to Furlough's et al. \cite{furlough2021attributing} findings where participants assigned nearly equivalent blame to robots as to environmental factors when the robot was portrayed as non-autonomous. Correspondingly, when the robot was depicted as autonomous, participants assigned almost as much blame to the robot as they did to the human counterpart. 
\newline In HHI, attribution of responsibility seems to be connected to the relationship’s intimacy between two collaborators \cite{sedikides1998self}. This phenomenon is assumed to be extendable to Human-Agent collaborations \cite{reeves1996media,matsui2021blame}.
Nass and Moon \cite{nass2000machines} observed that individuals instinctively engage in social processes and generate social responses when interacting with computers, even when they are not consciously aware of this behavior \footnote[7]{This is also refered to as the CASA (Computers are social actors) paradigm.}. These processes encompass various behavioral aspects such as gender stereotyping, distinguishing between "self" and "other," and expressions of politeness.
\newline In HRI, humans reported favoring robots that blamed themselves after mistakes occurred, as opposed to blaming the person or the team \cite{groom2010critic}. Participants’ perceptions and responses towards robots could be enhanced when the latter used politeness strategies \cite{torrey2009robots,takayama2009m}. 
This, however, could become a problem if the robot fits into other stereotypes which it then reinforces, such as being gendered as female. For instance, adolescents reported a decline in likability for a female robot expressing negative emotions, while conversely, they demonstrated an increased liking for a male robot displaying identical emotions \cite{calvo2020effects}.
Male participants reported liking male robots but not female robots when they rejected morally questionable commands \cite{jackson2020exploring}. This suggests that similar to societal patterns with women, female robots are less appreciated when they exhibit non-compliant or non-consensual behavior \cite{perugia2023robot}. 
While these preferences may exist, they reflect discriminatory practices in societal structures which can be used to leverage positive perceptions of technology, but perhaps should not be used, since they could reinforce gender stereotypes. 
Agency also matters in Tollon’s \cite{tollon2023responsibility} analysis, which suggests that emotional responses and responsibility attribution should be higher for AI that is perceived as having a more agent-like character. 
Tollon examines how reactive attitudes toward AI systems might impact our habits of attributing responsibility and questions whether maintaining an objective stance toward AI agents is reasonable. He contends that adopting such an attitude aligns with three specific aspects of responsibility—answerability, attributability, and accountability.\footnote[8]{This distinction is well addressed within the discussions of moral responsibility in the corresponding philosophical literature \cite{shoemaker2015responsibility,watson2004agency}.} Consequently, Tollon argues that AI systems do not erode our ability to attribute responsibility in these dimensions.

\section{Revisiting Reactive Attitudes}


The consensus in the responsibility literature is that P.F. Strawson´s “Freedom and Resentment” (1962, 2008) is one of the most influential papers of the debate. Its main force lies in the fact that it caused an impactful shift in the discussions around responsibility by responding to the debate about the compatibility of free will and responsibility. Although Strawson´s main aim was to reconcile the two opposing sides of the debate, namely the compatibilists and incompatibilists, he actually signaled a new tradition in the responsibility literature introducing what he named as the “reactive attitudes.”
\newline Thus, the strawsonian paradigm has shifted the debate of responsibility from the metaphysical question of whether we have free will to the examination of human relationships. This examination would impose that philosophers would take a closer look at how people are treated and perceived within different types of relationships. Although at first glance this perspective could be a useful tool for examining how vulnerable groups are often mistreated and overly exposed to negative reactive attitudes, this potential aspect of the strawsonian theory has been underexamined. 
Recent criticism has shed light on how the strawsonian tradition has overlooked specific assumptions that Strawson had to presuppose in order to introduce the paradigm of the reactive attitudes. For example, Ciurria \cite{ciurria2019intersectional,ciurria2023responsibility} has highlighted how much of the strawsonian understanding of interpersonal relationships is based on an ideal understanding of the social reality. 
Similarly, Manne \cite{manne2017down} underlines that often reactive attitudes track power imbalances and police vulnerable groups into specific social behaviors instead of tracking moral judgments.
\newline It seems that the strawsonian tradition underrepresented the fact that in human relationships we often confuse moral norms with social ones.  For example, in the classic book “The Adventures of Huckleberry Finn” (1876), the protagonist "Huck" is depicted struggling with what he thinks he should do according to the norms and regulations of society and what he thinks is the right thing to do \cite{arpaly1999praise}. Huck´s friend, Jim, is a slave who escaped his master. The morality of the time dictates that Huck takes Jim back to his master. But Huck feels that turning his friend in is wrong despite the moral rule.
Similarly, every example of civil disobedience, for instance, the famous cases of Rosa Parks and Mahatma Gandhi, is an example of how sometimes what we think are the demands of morality clash with actual social norms, and this is a two-way street.
Of course, it can be a challenge to discern moral from social norms as they seem to be intertwined. However, these norms can be different and sometimes, they can pull in opposite directions.
\newline Moral philosophy could have provided tools for vulnerable groups to navigate their reality, but instead, the problematic assumption that social norms are dictating what is moral is deeply rooted within most societies and continues to reinforce problematic narratives and phenomena such as victim blaming \cite{leventi2024victim}. 
Unfortunately, injustice and power imbalances are embedded within social norms, and if moral rules are informed by problematic social ones, then the responsibility judgments are open to the same problems. It could be assumed that moral philosophy presupposes an ideal world where social and moral norms coincide. However, at least in Strawson´s paper, there is no indication that he presupposes an ideal world state, where the social and moral norms are identical. 
The fact that we do not live in an ideal world cannot go unnoticed and not be addressed in moral philosophy. People who traditionally worked on moral issues were indeed representatives of groups that did not have to be challenged with issues such as victim blame, but steps have been made, so philosophy is more inclusive to more people who belong to different social groups. Consequently, the trajectory of research should be reevaluated in order to fit and acknowledge people who may come from different backgrounds and have different experiences in life. 
The reactive attitudes cannot work to support our practices if they just map social rules instead of moral. They will reinforce problematic behaviors \cite{ciurria2023responsibility} and they exclude people who do not fit in that ideal narrative. So if people do not respond in the right way to your concern, then you must be the problem. If someone, for example, steps on your foot and does not apologize, as Strawson expects that such a wrongdoer would feel obliged to do, then something is wrong with how you are viewed within the morality system. You are someone who can be disrespected. If a harmful event goes unacknowledged, it weighs down the victim\cite{hieronymi2001articulating}. Thus, reactive attitudes can fail to provide support to vulnerable groups. Arguably, Strawson does not seem to aim to create a comprehensive theory of moral responsibility that has all the narratives that are important for vulnerable groups. This can be too demanding for a theory to achieve in a single paper.\footnote[9]{Especially one written a few decades ago.} However, the discussions that span out of this theoretical framework show that there are specific themes that people working in this debate are and were interested in. The perspectives that are represented showcase that some voices are heard more than others within the realm of moral philosophy. For example, talking about the victim's well-being is not often seen as a good enough reason to blame a perpetrator with questionable normative competence. For opposing views see \cite{leventi2024perspective}.
\newline It is important to acknowledge that the idea of reactive attitudes will not work for everyone. There is a specific type of person that can demand compensation from others and these others will acknowledge their wrongdoing and then take steps to make amends. This tendency in the philosophical literature to assume ideal circumstances, where all agents are viewed as equals, without explicitly stating that as a presupposition can be deeply problematic. When a victim does not comply to this narrative, then their experiences are disregarded, not viewed as data and eventually erased \cite{d2023data}.
This situation of unaddressed assumptions can come across in philosophical discussions for many reasons, for example, philosophers might not be aware that they do not live in an ideal world, or they might think that an ideal world assumption is needed in order to understand and examine philosophical concepts.

\section{Social Identity, Norms and Power in HRI}

Social norms facilitate mutual understanding, cooperation and can even help empathizing with and protecting victims \cite{dunn2004politics, mead1925genesis}. In this section however we will focus on the less positive impacts of norms.
\newline Oppression consists of multiple dimensions, as is suggested by the notion of intersectionality, but for the purposes of the current paper, we will in this work focus on gender as an identity marker and indicator of power and social norms. Here, we conceptualize gender as Perugia and Lisy \cite{perugia2023robot} describe it, namely as a cognitive and operational framework, detached from inherent traits or bodily features. It functions as a structuring principle that is ingrained in social frameworks, performance, design, and norms. Since gender is commonly built into AI systems as a binary construct, our analysis is mainly going to orient itself in these terms, too (i.e. mostly focusing on the female/male division). Nevertheless, we are aware that there is a whole spectrum of human gender identities and we hope that in the future there will be more examples considering those in robot design.
Marking gender in AI has been a long practice, where e.g. mathematics was seen as a male field of expertise, and at the same time, as \emph{the} important form of intelligence \cite{adam2006artificial}. To establish what can be regarded as intelligent, familiarity played an important part; these familiarities might seem universal, however, they are based on specific individuals’ perspectives \cite{rhee2018robotic}. Thus, early AI researchers’ embodied experiences were presented as given, leading to stereotypical (mis-)representations, simplifications, and hierarchies. These depictions were additionally normatively charged. In this context, scripts are also relevant to how interactions play out. Scripts can be described as predefined sequences from familiar scenarios, which influence how actors are imagined to behave \cite{shank1977scripts}. Inscriptions in AI are therefore not neutral but privilege some subjects’ perspectives and erase others. These scripts create social norms which can easily be confused for moral norms in navigating through interpersonal relationships and interactions.
\newline Robots, by being prone to error and awkwardness, seem to make it necessary for humans to perform labor -sometimes invisible labor- to keep interactions intact. For instance, Pelikan, Reeves and Cantarutti \cite{pelikan2024encountering} showed that, while interactions with robots tend to be tested in a deliberate lab-based scenario, delivery robots in public spaces often require spontaneous 'accommodation work' by construction and service workers and other human passerbys.
Rhee \cite{rhee2018robotic} points out that particularly female labor is often connected with care activities and expectations. A fictional example of this assumption is Samantha from the movie Her, while in real life, Apple’s Siri and Amazon’s Alexa can be seen as exemplifications of this observation. The role of these personified helpers is to fulfill the needs of emotional support and relationship partners such as that of a spouse or parent, a role which is at the same time devalued in status and perceived worth. On the other hand, Male AIs, like Watson, IBM’s expert-systems AI personify, how Rhee formulates it, “models of authoritative expertise often ascribed to men” (\cite{rhee2018robotic} p.36), which usually are depicted to hold and spread knowledge but not to fulfill specific socially-supportive roles. 
\newline Erroneous, or as Strengers and Kennedy \cite{strengers2021smart} call it, glitchy behavior in AI systems will often be explained using stereotypes about women. For instance, Alexa has been described by journalists as “manic”, “hysterical” or “disturbing", while for instance the Halo version of Cortana was said to be in “bitch mode” or having “AI PMS” (\cite{strengers2021smart} p.148).
While AI has the potential of subverting social conventions and roles, these commercially used systems are rather reinforcing norms and stereotypes, to “keep women in their place. In so doing, robots and women are simultaneously represented as a ‘threat that must be controlled’” (\cite{strengers2021smart} p.148).
One of the factors that seems to influence how we hold different expectations and emotions towards different groups of people is attributions of power. To examine this idea, we will take a closer look at the fundamental principle of Datafeminism which investigates relationships of power. Klein and D’Ignizo \cite{d2023data} discuss the matrix of domination, which consists of structural, disciplinary, hegemonic, and interpersonal domains. The disciplinary realm exhibits a tendency to shift responsibility, neglect proper investigation, and engage in victim blaming while benefiting from an absence of consequences within the structural domain. This in turn would not be accepted if not for the support of the hegemonic domain (sphere of media and culture). This domain portrays men as powerful and women as submissive, trans people as breaking substantial norms and erasing nonbinary people. 
This means that governmental institutions and media channels contribute significantly to harmful but pervasive phenomena such as victim blaming.\footnote[10]{Within the scope of this paper we accept this statement is true. For more see \cite{taylor2020women,leventi2024victim}.} Researchers, as well as designers, and companies building AI systems and Robots therefore carry a risk of being part of the hegemonic domain and perpetuating these norms. While the matrix of domination has been replaced with more accurate models throughout the past, it is still a helpful framework to understand different sources of influence; these different dimensions can also blend into each other, e.g. when it comes to robot blame, both he interpersonal and the hegemonic sphere seem to play an important part.
\newline Klein and D’Ignizo furthermore encourage readers to analyze AI projects in regards to whose aims are considered, and whose are not. It might be efficient to use an apologetic strategy for the robot, suggesting that the robot takes all responsibility and blame. However, adopting this framework raises the question of what the long-term consequences will be, and whether this strategy prioritizes specific people’s goals and convenience. 
Considering gender stereotypes in service robots, Hu et al. \cite{hu2022effect} found that matching occupational gender stereotypes and robot gender increased participants’ willingness to engage but when errors occurred, this effect was reversed. It is important to underline that participant gender has also been shown to be a factor in how humans interact with artificial agents \cite{crowelly2009gendered}. Prior research has suggested that the perception of a robot is a consequence of an interaction of robot gendering, which interacts with gender and observant gender (if the latter two are not the same person) \cite{jackson2020exploring}. 
Lei and Rau \cite{lei2021should} found that female participants assigned less blame and more credit to the robot in a team task.
This supports the idea that interactions with robots happen within a context of power structures and narratives around social roles as described by Klein and D’Ignozio \cite{d2023data}.
At the same time, several studies discovered that the gendering of robots did not result in improved perception of the robot \cite{rea2015check}. For instance, Bryant, Borenstein, and Howard \cite{bryant2020should} did not find a correlation between trust and fulfillment of occupational gender stereotypes in robots. In addition, following gender norms does not have to be the most efficient and beneficial way: in educational settings, it seemed like a mismatch between robot genderedness and stereotypical tasks was even preferred \cite{reich2017ir}.  
\newline Nevertheless, alternative studies have revealed noteworthy distinctions in the ratings of male and female-gendered robots, particularly in areas that confirm common stereotypes such as emotional intelligence (rated higher for female robots) and agency (rated higher for male robots) \cite{chita2019gender,eyssel2012s}. These studies support the previously mentioned assumption that gender stereotypes and expectations of behavior extend from human-human to human-machine interaction \cite{nass1997machines}. These effects have been observed in interaction with conversational agents too, where gender bias and assumptions of ‘appropriate’ behavior translated into human-machine interactions \cite{curry2020conversational,moradbakhti2022men}. At the same time, behaviors that oppose stereotypical representations have been demonstrated to weaken gender-related presumptions \cite{olsson2018does}. This is in accordance with results by Erel, Carsenti, and Zuckerman \cite{erel2022carryover}, who found that experiences from HRI could carry over into subsequent HHI; thus interactions with robots have the potential to impact our interactions with other people both negatively and positively.


\section{Are Robots Responsible?}
\subsection{Robots' Influence on Blame Practices}

Babushkina \cite{babushkina2020robots} argues that robots in the case of failures serve merely as an empty placeholder where the person that actually carries responsibility is removed. Thus, reactive attitudes directed towards robots, i.e. blaming them, place a risk on our moral practices. This observation may be valid, however, we argue that it does not matter whether or not users should direct blame toward the robot or not; social behaviors are intuitive and do usually not occur as a result of long reflection, hence users risk extending this behavior onto robots, regardless if it is morally acceptable or not. As our blaming practices seem to be influenced by interpersonal and structural factors, similarly it can be examined whether identity factors within hierarchical power structures influence blame (mis-)directed responses towards the robots, interactants, or roboticists. Babushkina writes that the “root of the problem seems to be the perception of the robots’ responsibilities: it systematically escapes our attention that robots have quasi-responsibilities at best, and they do not free humans from blame” (\cite{babushkina2020robots} p.313). What seems to be missing here is the consideration of whether certain humans might be freed more or less from blame in interaction with quasi-responsibility-holding robots. Moreover, it seems that robots are vessels for not only social norms, but also moral norms; if people do not act in a way that works well together with robots, they may be blamed and framed in a way like civil disobedience is being framed. For example, if it seems that the robot requires help and the person who is cooperating with that robot is a woman then the stereotypical accepted behaviors would be presupposed. Namely, women are helpers and caregivers, so the woman should help the robot in need. On the other hand, the man cooperating with the robot in need would not be held responsible for not helping the robot, but he would be praised if he did as he would be performing a supererogatory act. 
\newline We agree with Babushkina in that blaming robots could pose a threat to our moral practices since the confusion of social with moral norms could become augmented through interactions with robots. Blaming the robot, i.e. an entity that cannot be held responsible, the people that should actually be held responsible are harder to pinpoint: “Scapegoating is unmerited, unwarranted blame. In psychological terms, it is understood as the redirection of your reactive attitudes from the offender to an entity innocent of the transgression causing those sentiments. As such, scapegoating is a moral transgression because it knowingly misplaces and misuses blame. A robotic assistant acts as a mediator to our relationships with each other and makes human contributions to harm hard to trace.”(\cite{babushkina2020robots} p.314).
Thus, co-liberation and justice (as described in \cite{d2023data}) can be harder to achieve for marginalized groups. In many cases of errors, neither the robot, nor the persons interacting with or being affect by the robot's actions should be blamed, but rather the manufacturer or robot designer. This connection however risks to get lost due to the robot presenting as its own entity. 
\newline Social robots, as we imagine and design them currently, contain a risk of creating a new form of victim-blaming; one where people who do not put in enough care labor but whose identity inscribes such responsibilities, do not do 'their job', and fulfill their duties and therefore can be blamed. 
By changing robot identities, we also change normative expectations of the interactants depending on their own identity. For instance, a childlike robot might elicit different expectations of a perceived female person compared to a perceived male person. When people interact with robots, they have to relate themselves to them and do it automatically, just as the people observing would do.
Robots are not just empty placeholders for blame; through their embodiment shaping the perceived nature (as we learn from Strengers and Kennedy \cite{strengers2021smart}, as well as Rhee \cite{rhee2018robotic}), robots are anything but empty. On the contrary, robots carry lots of projections, narratives, imaginaries on and within their bodies. This idea is supported by the findings of Erel et al \cite{erel2022carryover} which suggest that encounters with them influence our interactions with persons completely unrelated to the root of an event. 
These considerations however also show that it is not enough to contemplate whether and how blame is attributed to different actors in and as a consequence of Human-Robot interactions, but to concretely test how humans react depending on the embodiment and presumed social identity of the robot. 
In this case of course, there is a risk that the nuances which are considered in the Strawsonian paradigm (like the distinction between moral and social blame, for instance with the notion of resentment) would disappear through the reductive nature of empirical research. However, we believe that this worry itself could be an important contribution to HRI research by emphasizing the need of carefully describing and conceptualizing what it is that should be measured when participants react to the robot. In the following, we will explain how the Strawsonian paradigm aligns with and may even benefit from interventions within AI practices.

\subsection{Embracing the Glitch}

While offering a structured approach to analyzing interpersonal relationships with respect to responsibility, recent critiques have brought attention to certain shortcomings within the Strawsonian tradition. As mentioned previously, it has been pointed out, that the tradition tends to rely on an idealized perception of social reality \cite{ciurria2019intersectional,ciurria2023responsibility}.
Thus, reactive attitudes share a commonality with robots and AI systems, which are also built with an 'ideal' picture of reality, society, and human behavior in mind. One example of the latter is the concept of the "uncanny valley", which is a theory used to explain why certain robots seem unsettling to us; embedded within this theory are however assumptions of what or who can be viewed as a person and what does not belong in that category, by relying mostly on what is strange or not familiar to oneself and by for instance, potentially excluding persons with disabilities \cite{rhee2018robotic}. 
\newline One non-ideal factor is failure induced by a human collaborator or operator, a factor that is often neglected in HRI. Somasundaram et al. \cite{somasundaram2023intelligent} therefore developed a framework based on the notion of Intelligent Disobedience (ID), i.e. acting correctly when given orders are faulty or otherwise problematic.
That way, a robot can use supposed disruptions to improve situational outcomes. Furthermore, Winkle et al. \cite{winkle2022norm} have used robotic behavior to challenge gender norms and stereotypes, by letting a (as female gendered) conversational robot talk back -either with a rational explanation or attacking back- after being faced with abusive remarks. This study presents a valuable example of how the questionable depiction of robots or AI systems (thinking back to Alexa here), could be counteracted.
Although it might appear that there is a contradiction between the idea that gendered robots are perpetuating power imbalances and the suggestion of "accepting the glitch", and to use gendered robots for norm-challenging behavior, we believe that they can work parallel to each other. Understanding the nuance of our practices  and thus being able to alleviate the possible issues, demands that we are able to point out to such behaviors. Accepting or recognizing that gendering robots reinforces the state of structural injustice does not explain to the researcher the process or the extent of this phenomenon. It seems that this kind of understanding has been adopted by Winkle et al. \cite{winkle2022norm}.\footnote[11]{We thank Ingar Brinck for bringing this worry to our attention. }
In addition, Treusch's \cite{treusch2020robotic} use of robots for collaborative knitting challenges the prevailing focus on optimization and efficiency, which is especially common in industrial and service contexts. Instead, they concentrate on its disruptive and experiential effects.
As a strategy to repair the interaction after an error occurs, roboticists have for instance implemented apology- and politeness-behaviors \cite{mahmood2022owning, lee2010gracefully}. In these models, the context is however commonly excluded, as well as the actual cause of error, or how the robot was treated by the user.
Instead of, as Babushkina \cite{babushkina2020robots} states, merely posing a risk to our moral practices, robot blaming makes it apparent that these practices we have are flawed. There seems to be an underlined assumption that the agents that carry characteristics of being caregivers or are prescribed as providing some type of support are acceptable targets of negative reactive attitudes when they violate these duties. 
\newline Therefore, it seems like robotic errors should not be swept under the carpet, but rather make us question why these errors bother us in a particular way, namely as agents whose observers expect to produce the kind of labor that avoids errors and we blame them if they fail.

\section{Conclusion}

We have argued that reactive attitudes can play an important role in how we understand interactions between humans and robots. It is evident that if reactive attitudes track power imbalances while attributing and signaling blame to agents then this process will be translated to human-robot cooperation. 
A possible solution to this issue is to focus on robot designs that resemble human identities less and actually allow for sociality that avoids anthropomorphism. Instead, designers could use the concept of sociomorphing, a concept which has been suggested by Seibt et al. \cite{seibt2020sociomorphing} and is used to enable relationality and interactions with entities that are clearly not human-like but still capable of establishing some sort of social contact.
Although building avatars and robots that remind us of certain human groups seems to be what most designers have chosen, it could be argued that constitutes taking a (harmful) shortcut. In the long term, this adoption of human-like characteristics perpetuates already existing stereotypes. It also presumes that there is such a thing as the ideal 'universal' human and creates norms around what counts as labor and as appropriate social behavior \cite{rhee2018robotic}. Mitigation strategies, such as apology and politeness behaviors, should be applied cautiously, as they might influence social norms and expectations towards certain groups.
\newline We suggest that in addition to interdisciplinary reflective thinking in robotic design, there is a need for empirical research, both quantitative and qualitative, to investigate the consequences of blaming practices and identities ascribed to the robot that result from effortless transferring of human interaction scripts to HRI. Furthermore, we want to encourage more researchers to use HRI as a tool to challenge problematic societal practices by letting the robot take over some of the subversive labor of changing norms and expectations that are based on discriminatory practices.

\bibliographystyle{vancouver.bst}
\bibliography{references}

\end{document}